\documentclass[12pt]{article}
\usepackage{amsfonts,amsmath,amssymb,epsf}
\usepackage{amsmath}
\usepackage{graphicx,color}
\newcommand{\de}{\partial}
\newcommand{\be}{\begin{equation}}
\newcommand{\ba}{\begin{eqnarray}}
\newcommand{\ea}{\end{eqnarray}}
\newcommand{\ee}{\end{equation}}

\newcommand{\lr}{\leftrightarrow}
\newcommand{\f}{\frac}
\newcommand{\s}{\sqrt}

\newcommand{\ti}{\tilde}

\newcommand{\ddd}{\cdot\cdot\cdot}
\newcommand{\no}{\nonumber \\}
\newcommand{\la}{\langle}
\newcommand{\lb}{\rangle}
\newcommand{\bea}{\begin{eqnarray}}
\newcommand{\eea}{\end{eqnarray}}
\newcommand{\bes}{\begin{equation*}}
\newcommand{\beas}{\begin{eqnarray*}}
\newcommand{\eeas}{\end{eqnarray*}}
\newcommand{\bas}{\begin{array*}}
\newcommand{\eas}{\end{array*}}
\newcommand{\ees}{\end{equation*}}

\newcommand{\ep}{\epsilon}

\begin{document}

\begin{titlepage}
\thispagestyle{empty}

\begin{flushright}
YITP-15-16
\\
IPMU15-0024
\\
\end{flushright}


\begin{center}
\noindent{{\large \textbf{Surface/State Correspondence as a Generalized Holography}}}\\
\vspace{2cm}
Masamichi Miyaji $^a$
and
Tadashi Takayanagi $^{a,b}$
\vspace{1cm}

{\it
$^{a}$Yukawa Institute for Theoretical Physics,
Kyoto University, Kyoto 606-8502, Japan\\
$^{b}$Kavli Institute for the Physics and Mathematics of the Universe,\\
University of Tokyo, Kashiwa, Chiba 277-8582, Japan\\
}

\vskip 2em
\end{center}

\begin{abstract}
We propose a new duality relation between codimension two space-like surfaces in gravitational theories and quantum states in dual Hilbert spaces. This surface/state correspondence largely generalizes the idea of holography such that we do not need to rely on any existence of boundaries in gravitational spacetimes. The present idea is motivated by the recent interpretation of AdS/CFT in terms of the tensor networks so called MERA.  Moreover, we study this correspondence from the viewpoint of entanglement entropy and information metric. The Cramer-Rao bound in quantum estimation theory implies that the quantum fluctuations of radial coordinate of the AdS is highly suppressed in the large $N$ limit.
\end{abstract}

\end{titlepage}

\newpage

\section{Introduction}

Recent progresses in string theory strongly suggest that the idea of holography \cite{Hol}
will play a crucial role to construct a complete theory of quantum gravity. The holographic principle argues that gravitational theories are equivalent to non-gravitational theories which are defined as quantum many-body systems or quantum field theories. As in the AdS/CFT correspondence \cite{Maldacena}, which is the best known example of holography, a dual non-gravitational theory lives on the boundary of its original gravitational spacetime.

In the AdS/CFT, a dual conformal field theory (CFT) lives on a time-like boundary of an anti de-Sitter space (AdS). Since the boundary includes the time direction, the dual CFT is a dynamical theory in a Lorentzian space. We can also perform a Wick rotation of both sides and this leads to a duality between a CFT  on an Euclidean space and a gravity on a hyperbolic space. As have been confirmed by numerous papers, the AdS/CFT works perfectly in both signatures.

On the other hands, if we turn to other spacetimes such as de Sitter spaces (dS), the idea of holography, which assumes a theory on the boundary, gets much more complicated and subtle. For example, in de Sitter spaces, only available boundaries are space-like ones. One possibility of its holography is known as the dS/CFT correspondence \cite{St}, which argues that gravitational theories on de Sitter spaces are dual to some Euclidean CFTs on their space-like boundaries.
However, the AdS/CFT already argues that Euclidean CFTs are dual to gravity on hyperbolic spaces
and thus we need to better understand how the dS/CFT works compared with the AdS/CFT. Moreover, if we perform a Wick rotation to find an Euclidean space, we obtain a sphere which has no boundaries. Therefore these motivate us to consider a generalization of holographic principle
without referring to dual theories on boundaries. For this purpose, we need a new framework of correspondence principle and the main aim of this paper is to propose one such possibility.

In this paper, we would like to propose a new framework of correspondence between structures of gravitational theories on any spacetimes and those of quantum states in quantum many-body systems. Especially, we argue that each codimension two convex surface in gravitational spacetime is dual to a certain quantum state, which we call surface/state correspondence. This leads to an understanding that gravitational spacetimes emerge from various distributions of quantum states. As a particular example, our formulation provides a refined structure of AdS/CFT correspondence.

Its has been conjectured in \cite{Swingle} (see also \cite{Qi} for a modified version of this conjecture) that the AdS/CFT can be interpreted as a real space renormalization flow called entanglement renormalization or MERA \cite{MERA}. Refer to e.g. \cite{MoV,SwC,MIH,HaMa,MVP,EvVi,FTV} for more progresses in this direction.
Our proposal in this paper is highly motivated by this and the continuum version (called cMERA \cite{cMERA}) of this conjecture, studied in \cite{NRT,MNRT,MRTW}. In the cMERA interpretation of the AdS/CFT, we can consider dual quantum states for each values of radial coordinate $z$ of the AdS, corresponding to each step of the real space renormalization flow. We would also like to refer to earlier interesting arguments \cite{Ra} and \cite{SuMa}, where mechanism of emergences of gravitational spacetimes have been discussed from the viewpoint of quantum entanglement.

This paper is organized as follows. In section two, we will present our main proposal of surface/state correspondence. In section three, we will extend our proposed correspondence to the analysis of information metric. We will also comment on an implication of quantum estimation theory in the final subsection. In section four, we will discuss several examples. In section five, we will summarize our conclusions and discuss future problems.

\section{Surface/State Correspondence Proposal}

Consider a gravity on an arbitrary $d+2$ dimensional spacetime $M_{d+2}$. Below we would like to present our proposal which describes the gravity on $M_{d+2}$ in terms of quantum many-body systems. We assume that this gravitational theory is approximated by the Einstein gravity coupled to various matter fields, such as supergravity theories.  However we expect that our proposal described below can be generalized to any gravitational theories by taking to into quantum effects appropriately.

\subsection{A Basic Principle of Surface/State Correspondence}

We start with a very large Hilbert space ${\cal H}_{tot}$, associated with the total spacetime $M_{d+2}$. Since ${\cal H}_{tot}$ is often given by an infinite dimensional Hilbert space of a (generically non-local) quantum field theory, it is useful to introduce a UV cut off $\ep$, interpreted as a lattice constant. For example, the Hilbert space is identified with that of a $d+1$ dimensional CFT if $M_{d+2}$ is given by the $d+2$ dimensional AdS spacetime following AdS/CFT. However, our construction is much more general and we do not need even time-like boundaries of $M_{d+2}$ as long as we can assume the presence of the total Hilbert space ${\cal H}_{tot}$.

Now we take a codimension two (i.e. $d$ dimensional) surface $\Sigma$ in $M_{d+2}$. In this paper, we always require that this surface $\Sigma$ is convex. Here we call a surface
convex if an extremal surface $\gamma$ which ends at an arbitrary chosen $d-1$ dimensional submanifold in $\Sigma$ is always included inside of the region\footnote{If $\Sigma$ is an open surface, then we define the region to be surrounded by $\Sigma$ and the extremal surface which connects the boundary
$\de \Sigma$.} surrounded by $\Sigma$. In other words, there always exists a $d+1$ dimensional space-like surface $N_{\Sigma}$ which ends on $\Sigma$ such that the extremal surface $\gamma$ is completely included in $N_{\Sigma}$. This condition becomes important to define the entanglement entropy in the next subsection.

\begin{figure}[ttt]
\centering
\includegraphics[width=10cm]{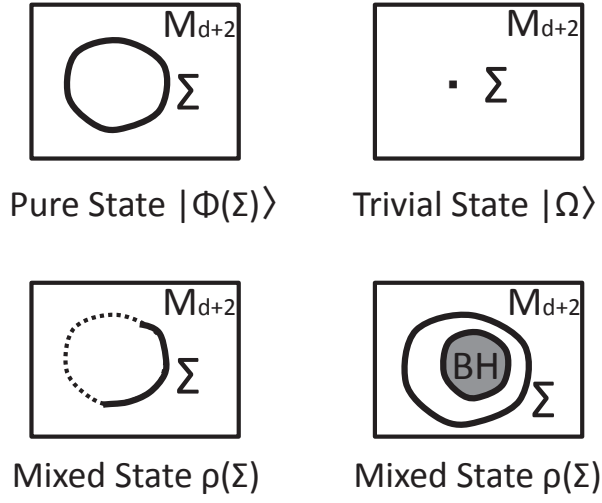}
\caption{Schematic Pictures of Surface/State Correspondence \label{fig:state}}
\end{figure}

First, let us focus on the case where $\Sigma$ is a closed convex surface which is topologically trivial i.e.  is homologous to a point. In this case, our basic principle starts by arguing that there exists a pure quantum state $|\Phi(\Sigma)\lb\in  {\cal H}_{tot}$ which corresponds to the surface $\Sigma$ (see the upper left picture in Fig.\ref{fig:state}):
\be
|\Phi(\Sigma)\lb\in  {\cal H}_{tot}\ \ \ \lr \ \ \ \Sigma \in M_{d+2} \ \ (\mbox{topologically trivial}). \label{pra}
\ee

More generally, if $\Sigma$ is a topologically non-trivial surface, then its corresponding state is given by a mixed state $\rho(\Sigma)$ in a Hilbert space
${\cal H}_{\Sigma}$, which is a subspace of ${\cal H}_{tot}$ (see the lower right picture in Fig.\ref{fig:state}):
 \be
\rho(\Sigma) \in  \mbox{End}({\cal H}_{\Sigma})\ \ \ \lr \ \ \ \Sigma \in M_{d+2} \ \ (\mbox{topologically non-trivial}). \label{praa}
\ee
 This is reduced to (\ref{pra}) if the surface is topologically trivial by setting $\rho(\Sigma)=|\Phi(\Sigma)\lb\la \Phi(\Sigma)|$.
This subspace ${\cal H}_{\Sigma}$ only depends on the topological class of $\Sigma$ and does not change under continuous deformation of $\Sigma$ in the sense of homology. In the case of the AdS eternal black hole \cite{MaE}, ${\cal H}_{tot}$ is given by a product of the two copies of CFT Hilbert space based on the thermofield construction. If $\Sigma$ is wrapped on the black hole horizon, then ${\cal H}_{\Sigma}$ is given by one of the two CFT Hilbert spaces. If $\Sigma$ is a topologically trivial surface, then we have ${\cal H}_{\Sigma}={\cal H}_{tot}$.

So far we assumed that $\Sigma$ is closed.
If $\Sigma$ has its boundary
$\de\Sigma$, the dual quantum state becomes a mixed state again as in (\ref{praa}), depicted in the lower left picture in Fig.\ref{fig:state}. When $\Sigma$ is a submanifold of a closed convex surface $\ti{\Sigma}$ in $M_{d+2}$, then its mixed state is given by tracing out the Hilbert space corresponding to the complement of $\Sigma$ in $\ti{\Sigma}$, denoted by ${\cal H}_{\ti{\Sigma}/\Sigma}$:
\be
\rho(\Sigma)=\mbox{Tr}_{{\cal H}_{\ti{\Sigma}/\Sigma}}[\rho(\ti{\Sigma})].
\ee

It is also useful to consider the zero size limit of a topologically
trivial surface $\Sigma$. We argue that it corresponds to the state $|\Omega\lb$ in ${\cal H}_{tot}$ with no real-space entanglement:
\be
\lim_{A(\Sigma)\to 0} |\Phi(\Sigma)\lb \to |\Omega\lb,  \label{trivial}
\ee
where $A(\Sigma)$ denotes the area of a topologically trivial manifold $\Sigma$, depicted in
the upper right picture in Fig.\ref{fig:state}.
This identification comes from an additional principle on the interpretation of the surface area as the sum of real-space entanglement, which will be explained in the next subsection.
In the recent paper \cite{MRTW}, such a state was identified with boundary states (or Cardy states) of a given CFT.

When two surfaces $\Sigma_1$ and $\Sigma_2$ are connected by a smooth deformation preserving convexity, we can describe this deformation by an integral of infinitesimal unitary transformations:
\ba
&& |\Phi(\Sigma_1)\lb=U(s_1,s_2) |\Phi(\Sigma_2)\lb,
\label{unita}  \\
&& U(s_1,s_2)\equiv P\cdot\exp\left[-i\int^{s_1}_{s_2} \hat{M}(s)ds\right],
\ea
where $P$ denotes the path-ordering and $\hat{M}(s)$ is a Hermitian operator; the parameter $s$ describes the continuous deformation such that $s=s_1$ and $s=s_2$ correspond to $\Sigma_1$ and $\Sigma_2$, respectively. In the expression (\ref{unita}) we assumed that the surfaces are topologically trivial so that they are dual to pure states.

When two surfaces $\Sigma_1$ and $\Sigma_2$ share the same boundaries
$\de \Sigma_1=\de \Sigma_2$ and are related to each other by a smooth deformation preserving convexity, the corresponding density matrices are related by unitary transformation
\be
\rho(\Sigma_1)=U(s_1,s_2)\rho(\Sigma_2)U^{-1}(s_1,s_2),\label{unitaa}
\ee
as long as there is no extremal surface between $\Sigma_1$ and $\Sigma_2$.
 This requirement of the absence of extremal surface is because of the requirement of convexity (for more details refer to the next subsection). The claim (\ref{unitaa}) can be naturally understood if we note that the deformation of $\Sigma_1$ with endpoints fixed acts non-trivially only for quantum entanglement inside $\Sigma_1$ and not for entanglement between $\Sigma_1$ and its complement.

Finally, let us comment that the precise definition of topological (non-)triviality is sometimes subtle. It is clearly defined if we can Wick-rotate $M_{d+2}$ into a regular Euclidean geometry by using the homology as in the case of holographic entanglement entropy \cite{HeTa}.
For example, in a typical example of black holes, the  non-trivial
surfaces correspond to the black hole horizons. However, such a Wick rotation is not always possible as in time-dependent black holes. We would like to refer to recent discussions \cite{HMRT,Head,HHMMR} and future developments in the context of holographic entanglement entropy for more details and we will not get into this subtle problem in this paper.

\subsection{Entanglement Entropy and Effective Entropy}

\begin{figure}[ttt]
\centering
\includegraphics[width=7cm]{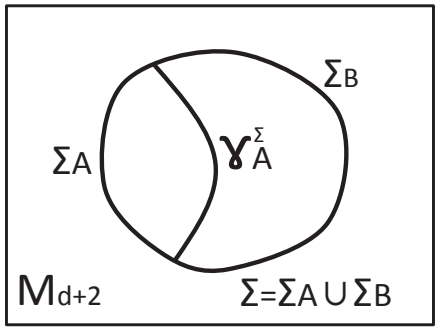}
\caption{Calculations of Entanglement Entropy in Surface/State Correspondence.
\label{fig:HEE}}
\end{figure}

A basic physical quantity which we can associate to a given (pure or mixed) quantum state for a $d$ dimensional (closed or open) surface $\Sigma$ is the entanglement entropy. For this purpose, we would like to consider a division of $\Sigma$ into two subregions: $\Sigma_A$ and $\Sigma_B$ such that $\Sigma=\Sigma_A\cup \Sigma_B$ and
$\Sigma_A\cap \Sigma_B=\phi$. Following the basic principle (\ref{praa}), the surface $\Sigma_A$ corresponds to a reduced density matrix $\rho_A$. The choice of subregions corresponds to a factorization of Hilbert space
\be
{\cal H}_{\Sigma}={\cal H}_A\otimes {\cal H}_B.
\ee
The reduced density matrix $\rho_A^\Sigma$ is defined by
\be
\rho_A^\Sigma=\mbox{Tr}_{{\cal H}_B}\rho(\Sigma).
\ee
Then we conjecture that the von-Neumann entropy of $\rho_A^\Sigma$ or equally the entanglement entropy $S_A^{\Sigma}$ (with respect to the quantum state $\rho(\Sigma)$)
is given by the area formula:
\be
S_A^\Sigma=\f{A(\gamma_A^{\Sigma})}{4G_N}, \label{eecon}
\ee
where $A(\gamma_A^{\Sigma})$ is the area of the extremal surface $\gamma_A^{\Sigma}$ in $M_{d+2}$ (refer to Fig.\ref{fig:HEE}). Also $G_N$ is the Newton constant of the $d+2$ dimensional
gravity on $M_{d+2}$. This extremal surface $\gamma_A^{\Sigma}$ is defined such that its boundary coincides with that of $\Sigma_A$ and that it is included in the region surrounded by $\Sigma$.  The latter condition requires that the surface $\Sigma$ should be convex as we mentioned.

In this formulation, the true dimension of the density matrix $\Sigma_A$ (or equally ${\cal H}_A$) is invariant under a smooth deformation with the two boundaries of $\Sigma_A$ fixed because they are related by the unitary transformation as in (\ref{unitaa}). Indeed, the von-Neumann entropy $S^{\Sigma}_A$ does not change under this deformation as is clear from (\ref{eecon}), which is consistent with the unitary evolution. Note that this unitary deformation of $\Sigma_A$ (denoted by $\hat{\Sigma}_A$) is terminated when it reaches the extremal surface $\gamma^\Sigma_A$. This is because we need to keep the closed surface $\hat{\Sigma}_A\cup \Sigma_B$ to be convex in order to define the reduced density matrix $\rho(\hat{\Sigma}_A)$. We can also argue that $\rho(\Sigma_A)$ does not change if we deform the surface $\Sigma_B$ with the same constraint.

Note that if we apply these claims to the AdS/CFT correspondence and take $\Sigma$ to be the AdS boundary, then (\ref{eecon}) is reduced to the holographic entanglement entropy formula \cite{RT}. Therefore our proposal (\ref{eecon}) can be regarded as a generalization of holographic entanglement entropy. For example, we can prove the strong subadditivity in the same way as that in the holographic entanglement entropy \cite{HeTa,AW}.\\

Now it is also intriguing to ask what is the quantum interpretation of the area of $\Sigma$ itself. Even though, $\Sigma$ is not an extremal surface in general, we can divide $\Sigma$ into infinitely many small subregions, which are all well-approximated by extremal surfaces. In such a small region, the geometry is approximated by a flat space and thus the extremal surfaces are given by flat planes. This consideration and the proposed correspondence (\ref{eecon}) lead to the following relation:
\be
\sum_{i}S_{A_i}^\Sigma=\f{A(\Sigma)}{4G_N}, \label{dimcon}
\ee
where $A_i$ describes the infinitesimally small portions of $\Sigma$ such that $\Sigma=\cup_i A_i$ and $A_i\cap A_j=\phi$. $S_{A_i}^\Sigma$ is the entanglement when we trace out the complement of $A_i$ inside $\Sigma$. It is useful to note that the left hand side of (\ref{dimcon}) is always larger than or equal to the total von-Neumann entropy for $\rho(\Sigma)$ owing to the subadditivity
relation.

We would like to call the left-hand side of (\ref{dimcon}) the effective entropy $S_{\mbox{eff}}(\Sigma)$. This quantity can be interpreted as the log of the effective dimension of the Hilbert space ${\cal H}_\Sigma$, written as $\log [\mbox{dim} {\cal H}^{\mbox{eff}}_{\Sigma}]$. Note that this effective dimension $\mbox{dim} {\cal H}^{\mbox{eff}}_{\Sigma}$ counts the dimension of effective degrees of freedom which participate in the entanglement between $A_i$ and its complement and is smaller than the actual dimension of the Hilbert space $\mbox{dim} {\cal H}_{\Sigma}$. We expect that this quantity $S_{\mbox{eff}}(\Sigma)$ is of the same order as the number of links in the tensor network representation which intersect with $\Sigma$. For this, refer to Fig.\ref{fig:TN} and the arguments in the next subsection.

Then the relation (\ref{dimcon}) can be written as\footnote{The areas of generic surfaces as in right-hand side of (\ref{effdim}) have recently been interpreted as an interesting quantity called differential entropy in the dual CFT \cite{BCCDH,MRS,Hu,Czech:2014tva}. Also another intriguing interpretation of the same quantity in terms of entanglement of gravitational theories has been conjectured in \cite{BiMy}. Note that our interpretation (\ref{effdim}) is apparently different from these in that our basic principle introduces quantum states for each surfaces.}
\be
S_{\mbox{eff}}(\Sigma)=\log [\mbox{dim} {\cal H}^{\mbox{eff}}_{\Sigma}]=\f{A(\Sigma)}{4G_N}.\label{effdim}
\ee
If we apply this to the AdS/CFT and choose $\Sigma$ to be the AdS boundary, then this estimation of effective dimension is reduced to the holographic bound \cite{SuWi}. Note also that the quantity (\ref{effdim}) does not depend on the details of the decomposition of $\Sigma$ into
the infinitesimally small pieces.

As a simplest example, consider the case where we shrink the closed surface $\Sigma$ to zero size. Then all $S_A$ and the effective dimension get vanishing. This means that the corresponding pure state $|\Phi(\Sigma)\lb$ does not have any real space entanglement, in spite of the fact that this state is defined in the infinitely large Hilbert space ${\cal H}_{tot}$ and this justifies our previous identification (\ref{trivial}).

If a closed or open surface $\Sigma$ is an extremal surface, then the entanglement entropy $S^\Sigma_A$ coincides with the effective entropy $S_{\mbox{eff}}(\Sigma)$ for the surface $\Sigma_A$. This means that the state $\rho(\Sigma_A)$
 saturates the subadditivity for any choices of subsystems in $\Sigma_A$. Therefore the density matrix $\rho(\Sigma_A)$ is a direct product of density matrices at each point:
 $\rho(\Sigma_A)=\otimes_{i}\rho(\Sigma_{A_i})$. This means that there is no true quantum entanglement within $\Sigma_A$. Indeed, a typical example of such a closed surface is the black hole horizon. If $\Sigma_A$ is an open surface as in Fig.\ref{fig:HEE}, $S^\Sigma_A$ and $S^\Sigma_{A_i}$ are non-trivial and these all come from the entanglement between $\Sigma_A$ and $\Sigma_B$ and not from the entanglement inside $\Sigma_A$.

\subsection{Relation to (c)MERA and Tensor Networks}

In the setup of AdS/CFT correspondence, our surface/state correspondence can be understood from the framework of real-space renormalization called multi-scale entanglement renormalization ansatz (MERA), introduced in \cite{MERA}. Our construction is closer to its continuum version called continuous MERA (cMERA) \cite{cMERA}. In the paper \cite{Swingle}, it has been argued that the mechanism of AdS/CFT can be understood as that of MERA if we consider discretizations (or lattice versions) of CFTs. Also, to realize continuum quantum field theoretic descriptions of AdS/CFT, we can relate cMERA to AdS/CFT \cite{NRT} to eliminates lattice artifacts. We would like to refer to theses references for detailed explanations of the conjectured equivalence between (c)MERA and AdS/CFT. Below we will give a brief summary of this argument.\footnote{For recent progresses on different approaches to construction of bulk theory from the boundary via RG flow, refer to
e.g. \cite{DMR,Lee,KLN,LPW,MiPo,BKM,Nak}.}

\begin{figure}[ttt]
\centering
\includegraphics[width=10cm]{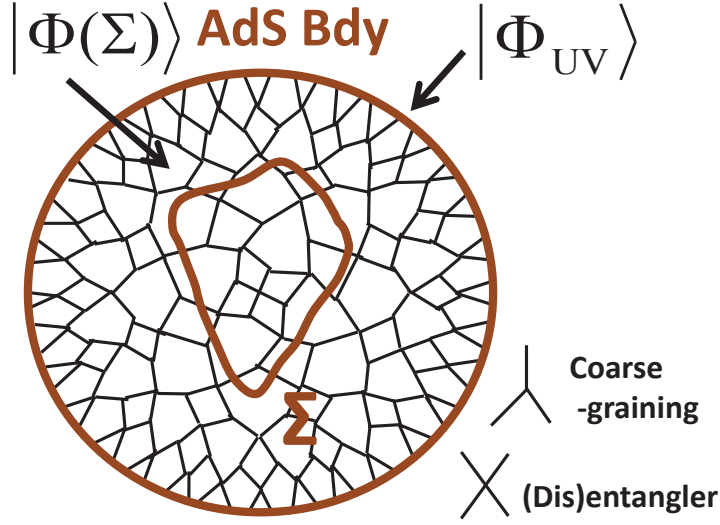}
\caption{Tensor Network of MERA and Surface/State Correspondence in AdS/CFT.
In the standard AdS/CFT correspondence, the CFT state (UV state) is given by $|\Phi_{UV}\lb$ and is defined by the quantum state realized at the boundary of the above MERA network. The black lines describe the flow of quantum states on discretized lattice points, which are acted by the coarse-graining and (dis)entangler operations. For a given convex closed surface $\Sigma$, we define the corresponding pure quantum state $|\Phi(\Sigma)\lb$ by contracting the indices of the tensors starting from the UV state $|\Phi_{UV}\lb$, following the tensor network as $\Sigma$ is homologous to the AdS boundary. Note that we can add a dummy trivial state $|0\lb$ for each coarse-graining operator so that both $|\Phi(\Sigma)\lb$ and $|\Phi_{UV}\lb$ live in the same Hilbert space ${\cal H}_{tot}$. Or equally, we can start from any point inside the region surrounded by $\Sigma$ and expand into $\Sigma$ to eventually find the state $|\Phi(\Sigma)\lb$. Therefore we can apply this correspondence to any networks even without boundaries.\label{fig:TN}}
\end{figure}

MERA gives a scheme of real space renormalization in terms of wave functions. This is rather different from the familiar Wilsonian approach, where the renormalization group flow in momentum space is studied in terms of effective actions. Suppose we are interested in the ground state of a quantum spin chain with a complicated Hamiltonian. We can coarse-grain the original spin system by combining two spins into a single spin according to an appropriate linear map (so called isometry). We can repeat this process arbitrary times until we reach just a single spin.
However, if we literally do this, short range entanglement of the coarse-grained quantum state is not neccesarily removed. Therefore, we need to cut out the short range entanglement just after each coarse-graining process by a unitary transformation, called a disentangler. These whole processes which modify the original spin state into a single spin state can be regarded as a network of spins, which is a particular example of tensor networks (see Fig.\ref{fig:TN}). The conjecture in \cite{Swingle} argues that the tensor network of MERA can be identified with the AdS spacetime, which can be qualitatively confirmed e.g. from evaluations of entanglement entropy.
For reviews on tensor networks refer to e.g. \cite{CVreview,Ereview,Oreview}.

In MERA or more generally tensor networks, we can pick up a convex closed surface in a given network and define a quantum state defined on the boundary of that part, by contracting indices of matrices of disentanglers and coarse-grainings (see the surface $\Sigma$ in Fig.\ref{fig:TN}). This consideration naturally leads to our surface/state correspondence assuming that the tensor networks are equivalent to gravitational theories. For this connection, as already mentioned in \cite{NRT} we do not need any presence of actual boundaries in a gravitational spacetime required usually in holography.

It is also useful to  note that the number of intersections between $\Sigma$ and the links in the tensor estimate the effective entropy $S(\Sigma)$. On the other hand, the entanglement entropy $S^{\Sigma}_A$ is estimated by the minimum number of intersections for curves which are homologous to $A$. However, we are not arguing the precise match for these estimations as each links in tensor networks are not necessarily maximally entangled.

In the cMERA formulation \cite{cMERA}, we focus on a one parameter family of quantum states $|\Psi(u)\lb$. The parameter $u$ $(-\infty<u<0)$ denotes the scale of renormalization such that the momentum cutoff is given by $|k|\leq e^u/\ep$, where $\ep$ is the original UV cut off (lattice spacing) of the theory we consider. The UV limit corresponds to $u=0$ and its state $|\Phi(0)\lb$ coincides with the vacuum of a given CFT.
On the other hand, $u=-\infty$ corresponds to the IR limit and with an appropriate IR
cutoff (i.e. a very small mass term) this IR state is given by the state $|\Omega\lb$ with no real space entanglement. We can identify such states with boundary states \cite{MRTW} in CFTs with an appropriate regularization.

Scince the vacuum state $|\Psi(0)\lb$ is highly entangled, we can construct this state by adding quantum entanglement to the trivial state $|\Omega\lb$ as
\ba\label{kmera}
|\Psi(u)\lb = Pe^{-i\int^u_{u_{IR}}(K(s)+L)ds}|\Omega\lb, \label{phfoww}
\ea
where $P$ denotes the path-ordering \cite{cMERA}. The hermitian operators $K(s)$ and $L$ describe the disentangler and the coarse-graining procedure. For our purpose it is useful to introduce another (equivalent) formulation \cite{NRT} given in terms of the state
\begin{align}
|\Phi(u)\lb &\equiv  e^{iuL}|\Psi(u)\lb
= Pe^{-i\int^u_{u_{IR}}\hat{K}(s)ds}|\Omega\lb,  \label{phfow}
\end{align}
where $\hat{K}(s)$ is defined by the disentangler in the interaction picture
\be
\hat{K}(u)=e^{iuL}K(u)e^{-iuL}.
\ee
Note that we perform the scale transformation after the disentangling for the states $|\Psi(u)\lb$, while we do not for $|\Phi(u)\lb$.

One important idea of cMERA is to keep the dimension of total Hilbert space to be the same. This can be realized by combining the coarse-graining procedure with adding trivial dummy states (e.g. $|0\lb$ at each lattice point) to keep the dimension same, so that the coarse-graining is described by a unitary transformation. We can do the same procedure for the coarse-graining operator in the network of Fig.\ref{fig:TN} so that the states $|\Phi(\Sigma)\lb$ live in the same Hilbert space ${\cal H}_{tot}$.  Therefore any states $|\Phi(u)\lb$ belong to states in the large Hilbert space ${\cal H}_{tot}$ defined in the UV theory. This trick was also had in mind in our discussions of section 2, where we argued that the Hilbert space does not change as long as the deformation of a surface preserves the homology class. For our purpose, we have in mind a generalization of cMERA such that we can choose the parameter $u$ for any inhomogeneous coarse-graining procedure (or inhomogeneous RG flow).

In the AdS/CFT viewpoint, this parameter $u$ corresponds to the coordinate of extra dimension defined by the following metric of Poincare AdS$_{d+2}$:
\be
ds^2=R^2 du^2+\f{R^2}{\ep^2}e^{2u}dx_\mu dx^\mu,\ \ \  \label{padsmetm}
\ee
where $R$ is the AdS radius and $\ep$ is the UV cut off of the dual CFT.

It is also possible to extend the cMERA formalism to finite temperature CFTs, which has been done in \cite{MNRT} for free scalar field theories. The MERA for quantum many-body systems at finite temperature was recently formulated in \cite{FTV} by improving the construction in \cite{Swingle,MIH}. The finite temperature (c)MERA can nicely be interpreted as the geometry of a double sided AdS black hole.

\section{Information Metric}

Consider two different closed surfaces $\Sigma$ and $\Sigma'$, which are both topologically trivial.　According to (\ref{pra}), we can associate them with pure states in the common Hilbert space ${\cal H}_{tot}$. This allows us to define the inner product $\la \Phi(\Sigma)|\Phi(\Sigma')\lb$ between them and provides us with additional information on the structure of quantum states. In this section we will focus on the inner product when the two surfaces $\Sigma$ and $\Sigma'$ are very close to each other by extracting the information metric and will study its dual gravity description. We will also comment on an implication of quantum estimation theory in the final subsection.

\subsection{Information Metric and Gravity Dual}

\begin{figure}[ttt]
\centering
\includegraphics[width=6cm]{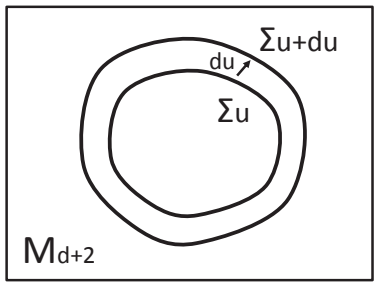}
\caption{Infinitesimal Deformations of Surfaces for Information Metric. \label{fig:IP}}
\end{figure}

Consider a one parameter family of codimension two closed surfaces $\Sigma_u$ and we assume that
the surface $\Sigma_u$ for any $u$  is topologically trivial such that it corresponds to the pure state $|\Phi(\Sigma_u)\lb$. We are interested in an infinitesimally shift $du$ of $u$ and in the inner product $\la\Phi(\Sigma_u)|\Phi(\Sigma_{u+du})\lb$ (refer to Fig.\ref{fig:IP}). Note that this inner product approaches to $1$ in the limit $du\to 0$ and it is straightforward to confirm that it always behaves
as $1-\la\Phi(\Sigma_u)|\Phi(\Sigma_{u+du})\lb=O(du^2)$. Therefore we will define the quantity
$G^{(B)}_{uu}$ as follows in the limit $du\to 0$
\be
1-|\la \Phi(\Sigma_u)|\Phi(\Sigma_{u+du})\lb|=G^{(B)}_{uu}du^2.  \label{infmetb}
\ee
This quantity $G^{(B)}_{uu}$ is called the Fisher information metric (in terms of Bures distance) \cite{Bookone,Booktwo,Bookthree,Book} and measures the distance between two different quantum states.\footnote{As we will mention in the final subsection in this section, there is another definition of Fisher information metric
$G^{(S)}_{uu}$ based on the relative entropy. These two metrics $G^{(B)}_{uu}$ and $G^{(S)}_{uu}$ are equivalent only for the classical states. However, in this paper we will not distinguish them seriously and assume that the results in this paper can be applied to for both metrics.}
We will give a more general definition of this metric in a later subsection including mixed states.

We describe the metric of $M_{d+2}$ by using the Gaussian normal coordinate as follows
\be
ds^2=R^2 du^2+g_{\mu\nu}(x,u)dx^\mu dx^\nu, \ \ \ (\mu,\nu=0,\ddd,d), \label{metref}
\ee
and consider the shift of $u$. Here $R$ is a constant with the length dimension so that $u$ becomes dimensionless. Our formulation is covariant on the choice of $R$.

In this setup, we conjecture that the information metric $G^{(B)}_{uu}$ defined by (\ref{infmetb}) is expressed in terms of the gravity on $M_{d+2}$ in the following form:
\be
G^{(B)}_{uu}=\f{1}{G_N}\int_{\Sigma_u} dx^d \s{g(x)}\int_{\Sigma_u}  dy^d \s{g(y)}\cdot P_{\mu\nu\xi\eta}(x,y,u)\left(\f{\de g^{\mu\nu}(x,u)}{\de u}\right)\left(\f{\de g^{\xi\eta}(y,u)}{\de u}\right), \label{Eincon}
\ee
where $G_N$ is the Newton constant of
the gravity on $M_{d+2}$ and $P_{\mu\nu\xi\eta}$ is a certain function, which is proportional to the amount of the degrees of freedom (such as the central charges in CFTs). Note that even though $x_\mu$ and $y_\mu$ are originally coordinates of $d+1$ dimensional space, we restricted to the $d$ dimensional space-like surface $\Sigma_u$ in the integrals in (\ref{Eincon}).

If we assume that the metric $g_{\mu\nu}(x,u)$ does not depend on the transverse coordinate
$x$, then we argue that (\ref{Eincon}) gets simplified into
\be
G^{(B)}_{uu}=\f{1}{G_N}\left(\int_{\Sigma_u} dx^d\s{g}\right)\cdot \hat{P}_{\mu\nu\xi\eta}(u)\left(\f{\de g^{\mu\nu}(u)}{\de u}\right)\left(\f{\de g^{\xi\eta}(u)}{\de u}\right). \label{Eincons}
\ee
 In order to match with cMERA results, as we will discuss shortly later,
 the tensor $\hat{P}_{\mu\nu\xi\eta}$ should scale like $g_{\mu\nu}g_{\xi\eta}$ under the Weyl scaling. The fact that the information metric is non-negative suggests that $\hat{P}$ is also non-negative. From this estimation we can find the tensor $P$ in (\ref{Eincon}) scales
 $P_{\mu\nu\xi\eta}(x,y,u)\sim \delta^d(x-y)g_{\mu\nu}g_{\xi\eta}/\s{g}+\ddd$, where we abbreviated non-local terms.

To understand the form (\ref{Eincon}), it is useful to think of an artificial Hamiltonian $H(u)$ such that the state $|\Phi(\Sigma_u)\lb$ is the ground state of $H(u)$ for each $u$. Note that in general the real time evolution by the time $t$ can be described by another true Hamiltonian $H_{ture}(u)$. Especially, in the
cMERA description of a CFT ground state, $H(u)$ is given by the Hamiltonian after the entanglement renormalization and coincides with the true Hamiltonian $H_{ture}(u)$ because the time evolution is trivial.

Then by using the standard second order perturbation theory of quantum mechanics (assuming no ground state degeneracy) we can derive
\be
1-|\la \Phi(\Sigma_u)|\Phi(\Sigma_{u+du})\lb|=(du)^2\cdot  \sum_{m\neq 0}\f{|\la m|\de_u H(u)|0\lb|^2}{(\Delta E_m)^2}, \label{genme}
\ee
where the ground state $|0\lb$ should be regarded as $|\Phi(\Sigma_u)\lb$ and $|m\lb$ denote all of its excited states; $\Delta E_m$ is the energy difference between $|m\lb$ and $|0\lb$ w.r.t
the Hamiltonian $H(u)$.  Since the infinitesimal change $\delta g_{\mu\nu}$ in (\ref{metref}) linearly affects the Hamiltonian $\delta H(u)=\int_{\Sigma_u} dx^d \delta g^{\mu\nu}(x) O_{\mu\nu}(x)$, where $O_{\mu\nu}$ is a certain operator which is analogous to the energy momentum tensor $T_{\mu\nu}$. Because for the infinitesimal change of $u$
we have $\delta g_{\mu\nu}=du\cdot \f{\de g_{\mu\nu}}{\de u}$, we reproduce the expression (\ref{Eincon}). Note also that (\ref{genme}) shows the well-known fact that the information metric $G^{(B)}_{uu}$ is non-negative for any unitary theories.

To estimate the normalization of $P$ and $\hat{P}$ in (\ref{Eincon}) and (\ref{Eincons}), consider the cMERA
description of a $d+1$ dimensional CFT ground state. The information metric in cMERA was computed in \cite{NRT} for surfaces $\Sigma_u$ defined by a fixed $u$ in the AdS space (\ref{padsmetm})
and the result is given by
\be
G^{(B)}_{uu}= N_{deg}\cdot \f{V_d}{\ep^d}e^{du}\sim S_{\mbox{eff}}(\Sigma_u),  \label{gbcft}
\ee
where $N_{deg}$ estimates the number of fields and is proportional the central charge. It is also useful to notice that the right-hand side of (\ref{gbcft}) is identical to
the effective entropy (\ref{effdim}) up to an order one factor.

Note that even though this behavior (\ref{gbcft}) 　was 　derived for a free massless scalar field theory, we naturally expect it can also be applicable to any CFTs due to the scaling property, where it is proportional to the effective volume of phase space. We can indeed confirm that (\ref{gbcft}) can be reproduced from (\ref{Eincons}) by substituting the metric of pure AdS (\ref{padsmetm}) and noting that $\hat{P}_{\mu\nu\xi\eta}\de_u g^{\mu\nu}\de_u g^{\xi\eta}$ is an $O(1)$ constant and that the ratio $R^d/G_N$ is proportional to the central charge of the CFT.

So far we only discussed the information metric for topologically trivial closed surfaces, which correspond to pure states. However, it is natural to expect that our expressions (\ref{Eincon}) and (\ref{Eincons}) can be applied to open surfaces and topologically non-trivial closed surfaces, which are dual to mixed states. Indeed, we can define the Fisher information metric for mixed states as we will explain in section 3.4.

Finally, we would like to suggest a plausible argument to fix the form of the tensor $\hat{P}$.
For simplicity let us assume that our space is static. Consider a AdS/CFT setup when $\Sigma_u$ becomes extremal (or equally minimal) at $u=u_0$. As we mention, the dual state is very special in that there is no real space entanglement inside $\Sigma_{u_0}$ as the subadditivity is saturated. As we move from the AdS boundary to the bulk, the information metric $G_{uu}$ is expected to decrease\footnote{This looks analogous to the c-theorem. We would like to come back to more details in \cite{IP}.} and eventually it becomes zero. However we know that $G_{uu}$ is non-negative and thus this flow is terminated at this point. Indeed, according to our arguments in previous section, we cannot move the surface across the extremal surface (see our argument near Fig.\ref{fig:HEE}). Therefore this implies that we have $G_{uu}=0$ at $u=u_0$.
 As the condition of extremal surface is given by the vanishing of the trace of extrinsic curvature: $K_u=g_{\bar{\mu}\bar{\nu}}\de_u g^{\bar{\mu}\bar{\nu}}=0$, we find the tensor $\hat{P}$ should only have the spacial components and is given by the form $\hat{P}_{\bar{\mu}\bar{\nu}\bar{\xi}\bar{\eta}}=C_P\cdot g_{\bar{\mu}\bar{\nu}}g_{\bar{\xi}\bar{\eta}}$,
where $(\bar{\mu},\bar{\nu},\bar{\xi},\bar{\eta})$ denotes the $d$ dimensional spacial part of the Lorentzian $d+1$ dimensional indices $(\mu,\nu,\xi,\eta)$. The coefficient $C_P$ is proportional to the degrees of freedom of the CFT. We can extend this argument to the time-dependent case as our present argument tells us that $G_{uu}$ is proportional to the integral $\f{1}{G_N}\int_{\Sigma_u}dx^d\s{g}K_u^2$, where $K_u$ is the trace of extrinsic curvature in the direction of our surface deformation.

\subsection{Analysis of Information Metric in AdS/CFT}

Here we would like to present an explicit analysis of information metric by concentrating on the AdS/CFT example by applying the cMERA construction. We focus on the state $|\Psi(u)\lb$ defined in (\ref{phfoww}) instead of our standard state $|\Phi(u)\lb$. Note that they are related by the scale transformation as in (\ref{phfow}). Thus the information metric we compute in this subsection is different from the original one (\ref{infmetb}). Nevertheless it may be still helpful to know the general behavior of information metric in CFTs.

Below we will evaluate the Fisher information metric for the state $|\Psi(u)\lb$. We would like to postpone general studies of Fisher information metric in quantum field theories to the work \cite{IP}. Note that the change of $u$ is simply interpreted as the standard renomalization group flow in the relativistic field theories. Refer also to \cite{Cad} for interesting results of the inner products under quantum quenches.

Consider the ground states $|0:V_{i}\lb$ $(i=1,2)$ for the theories whose Euclidean actions are given by $S_{CFT}+\int dt V_{i}(t)$, where $V_{i}(t)$ describes a infinitesimally small perturbation around a CFT defined by the action $S_{CFT}$; $t$ is now an Euclidean time. Then the overlap between these vacuum states are given by the following formula \cite{IP}
\be
\la 0:V_{2}|0:V_{1}\lb=\f{Z[\int dt \theta(-t)V_{1}(t)+\int dt\theta(t)V_{2}(t)]}{\sqrt{Z[\int dt V_{1}(t)]Z[\int dt V_{2}(t)]}},  \label{inpq}
\ee
where $Z[V]$ denotes the partition function in the presence of the interaction $V$.
If $V_{i}$s are infinitesimal, we can expand (\ref{inpq}) as
\be
\la 0:V_{1}|0:V_{2}\lb=1-\f{1}{2} \left\la \int^\infty_{\ep} dt (V_{1}(t)-V_{2}(t))
\int^{-\ep}_{-\infty} dt' (V_{1}(t')-V_{2}(t'))\right\lb+O(V^{3}),  \label{inghs}
\ee
where we assumed time reversal symmetry of the interaction $V_i(t)$. Note also that we introduced the UV cut off $\ep$ to remove the divergent contributions when the positions of
two interaction vertices collide. Moreover we find the inner product (\ref{inghs}) is real valued by considering the case where $V_i$ are a hermitian operators.

If we express the gravitational spacetime metric as in the Fefferman-Graham coordinate
\be
ds^2=R^2du^2+\f{R^2}{\ep^2}e^{2u}\hat{g}_{\mu\nu}(x,u)dx^\mu dx^\nu,\ \ \ (\mu,\nu=0,1,2,\ddd,d). \label{rrr}
\ee
where $R$ is the AdS radius and $\ep$ is the UV cut off of the dual CFT. The rescaled metric
$\hat{g}$  corresponds to the metric in the CFT.

We assume that the metric perturbation is time-independent so that the action for the Euclidean theory on $\Sigma_u$ is given by
\be
S_{CFT}+\int dt d^{d}x~ \delta\hat{g}_{\mu\nu}(x,u)T^{\mu\nu}(x,t), \label{qqw}
\ee
where we defined $\hat{g}_{\mu\nu}=\eta_{\mu\nu}+\delta\hat{g}_{\mu\nu}$.

In this setup, we can show\footnote{Strictly speaking, the deformation in (\ref{qqw}) corresponds to only the shift $\hat{g}_{\mu\nu}(x,u)\to \hat{g}_{\mu\nu}(x,u+du)$ with the warp factor $e^{2u}$ unchanged in the metric (\ref{rrr}). However, we can find the same result even if we do the infinitesimal shift $u$ in $e^{2u}$, owing to the fact that the scale invariance of our CFT leads to the identity $Q^\mu_{\mu\sigma\rho}=Q_{\mu\nu}{}^{\rho}_{\rho}=0$.}
\ba
&&1-|\la \Psi(u)|\Psi(u+du)\lb|
\no&&=\f{(du)^2}{2} \int d^{d}x d^{d}y \left(\f{\de \hat{g}^{\mu\nu}(x,u)}{\de u}\right)
\left(\f{\de\hat{g}^{\sigma\rho}(y,u)}{\de u}\right)Q_{\mu\nu\sigma\rho}(x-y),
\label{qqes}
\ea
where we introduced the tensor $Q$ as follows:
\be
Q_{\mu\nu\sigma\rho}(x-y)=\int^\infty_\ep dt\int^{-\ep}_{-\infty} dt'\la T_{\mu\nu}(x,t)T_{\sigma\rho}(y,t')\lb.
\ee
We can calculate this by using the explicit expression \cite{Os}:
\be
\la T_{\mu\nu}(x,t)T_{\sigma\rho}(y,t')\lb=\f{C_{T}}{((x-y)^2+(t-t')^2)^{d+1}}R_{\mu\alpha}(x,t:y,t')R_{\nu\beta}(x,t;y,t')
\mathcal{E}^{T}_{\alpha\beta:\sigma\rho},
\ee
where
\be
\mathcal{E}^{T}_{\mu\nu:\eta\xi}=\f{1}{2}(\delta_{\mu\eta}\delta_{\nu\xi}+\delta_{\mu\xi}\delta_{\nu\eta})-\f{1}{d+1}\delta_{\mu\nu}\delta_{\xi\eta}
\ee
\be
R_{\mu\nu}(x,t:y,t')=\delta_{\mu\nu}-2\f{(x-y)_{\mu}(x-y)_{\nu}}{(x-y)^2+(t-t')^2}.
\ee
Note that in even dimensional CFTs, the coefficient $C_T$ is proportional to the central charge of the dual CFT in the Einstein gravity description of AdS/CFT.

For simplicity, let us set $d=1$ i.e. a two dimensional CFT and compute the tensor $Q_{\mu\nu\sigma\rho}$ explicitly. To simplify the expressions we define the components $(a,b,c)$ so that they denote the pairs of components $(tt,xx,tx)$. Note that in this convention the tensor $Q$ is written as a symmetric $3\times 3$ matrix. We obtain
\ba
&&Q_{aa}=Q_{bb}=-Q_{ab}=-Q_{cc}=C_T\f{3\ep^4-6\ep^2(x-y)^2-(x-y)^4}{12\left((x-y)^2+\ep^2\right)^3}, \no
&&Q_{ac}=-Q_{bc}=C_T\f{2\ep^3(x-y)}{3\left((x-y)^2+\ep^2\right)^3}=-\f{\pi}{12}C_T\delta'(x-y),  \label{wwq} \no
\ea
where we took $\ep\to 0$ limit in the final expression.

It is also useful to perform the Fourier transformation:
$Q_{\mu\nu\rho\sigma}(x-y)=\int^\infty_{-\infty}dk Q_{\mu\nu\rho\sigma}(k)e^{-ikx}$, which leads to
\ba
&& Q_{aa}=Q_{bb}=-Q_{ab}=-Q_{cc}=\f{1}{24}C_T|k|,\no
&& Q_{ac}=-Q_{bc}=\f{1}{24}iC_T k.
\ea

After we goes back to the original Lorentz signature by the Wick rotation $t\to it$ we get
\ba
&& Q_{aa}=Q_{bb}=Q_{ab}=Q_{cc}=\f{1}{24}C_T|k|,\no
&& Q_{ac}=Q_{bc}=\f{1}{24} C_Tk.\label{wewq}
\ea
In particular, if we assume that the metric $\hat{g}$ does not depend on the coordinate $x^\mu$, we simply find $|\la \Psi(\Sigma_u)|\Psi(\Sigma_{u+du})\lb|=1$ i.e. the information metric is trivial as follows from (\ref{wwq}) and (\ref{wewq}).

In the presence of more general $x$-dependent metric perturbation, we can evaluate (\ref{qqes}) as follows:
\ba
&& 1-|\la \Psi(u)|\Psi(u+du)\lb| \no
&& =\f{\pi^2}{6} C_T (du)^2
\int^\infty_{-\infty}\! dk~ |k|\left(|\de_u\hat{g}^{tt}(k)|^2\!+\!2\de_u\hat{g}^{tt}(k)\de_u\hat{g}^{xx}(-k)\!
+\!|\de_u\hat{g}^{xx}(k)|^2\!+\!4|\de_u\hat{g}^{tx}(k)|^2\right).\no
\label{hatfis}
\ea
We can easily confirm that the information metric $G^{(B)}_{uu}$ is indeed non-negative.

 The relation between the CFT metric $\hat{g}_{\mu\nu}$ defined by (\ref{rrr}) and the one $g_{\mu\nu}$ in (\ref{metref}) is given by $g_{\mu\nu}=\f{R^2}{\ep^2}e^{2u}\hat{g}_{\mu\nu}$. Note also the familiar relation $C_T\propto R/G_N$ and remember that the UV cut off in the momentum integral in (\ref{hatfis}) will be $1/\ep$. Since the transformation from the information metric for $|\Psi(u)\lb$ to that for $|\Phi(u)\lb$ corresponds to removing the scale factor $e^{2u}$ in ($d$ dimensional) spacial metric components, it is natural to get the form (\ref{Eincon}) and (\ref{Eincons}) in the $|\Phi(u)\lb$ frame.

\subsection{Distance between Quantum Mixed States}

Here we would like to briefly review the information metric. First we should admit that the definition of information metric is not unique in quantum theory. For example, we can define two different Fisher information
metrics $G^{(B)}_{\theta\theta}$ and $G^{(S)}_{\theta\theta}$ as follows
(see e.g.\cite{Bookone,Booktwo,Bookthree}):
\ba
&& G^{(B)}_{\theta\theta}du^2
=B(\rho_{\theta+d\theta},\rho_{\theta}),\no
&& 2G^{(S)}_{\theta\theta}du^2=S(\rho_{\theta+d\theta}||\rho_{\theta}),  \label{distrh}
\ea
where we introduced two different measures of distances between two quantum states given by the density matrix
$\rho$ and $\sigma$
\ba
&& B(\rho,\sigma)=1-\mbox{Tr}[\s{\s{\rho}\sigma\s{\rho}}],\no
&& S(\rho||\sigma)=\mbox{Tr}[\rho(\log\rho-\log\sigma)].
\ea
The quantity $B(\rho,\sigma)$ is called the Bures distance,
while $S(\rho||\sigma)$ is known as the relative entropy, both of which measure entropic distance between two quantum states. Notice that in particular, for two pure states $|u\lb\la u|$ and
$|v\lb\la v|$ we can show
\be
B\left(|u\lb\la u|,|v\lb\la v|\right)=1-|\la u|v\lb|.
\ee

It is known that these two different metrics $G^{(B)}_{\theta\theta}$ and $G^{(S)}_{\theta\theta}$ coincide for classical states, while
they do not for generic quantum states. In this paper we simply focus on the former: $G^{(B)}_{\theta\theta}$ for our convenience assuming that the difference is not important for our purpose. We would also like to note that the distance (\ref{infmetb}) in our conjectured surface/state correspondence can be naturally generalized for that between two mixed states, corresponding to topologically non-trivial surfaces.

\subsection{Quantum Estimation Theory and Information Metric}

The information metric plays an important role in (quantum) estimation theory. Assume that the density matrix is parameterized by a real value $\theta\in $R, denoted by $\rho_{\theta}$. Now we would like to estimate the value of $\theta$ based on so called the POVM measurement. We define $X$ to be an operation which describes the measurement of $\theta$ such that
its expectation value $\la X\lb$ for $\rho_{\theta}$ coincides with $\theta$. Then we would like to estimate the mean square error $\la (\delta\theta)^2\lb=\la (X-\theta)^2 \lb$. In this setup, it has been known
that this error $\la (\delta\theta)^2\lb$ is bounded by the inverse of Fisher information metric as follows:
\be
\la (\delta \theta)^2\lb \geq \f{1}{8G^{(B)}_{\theta\theta}}. \label{cro}
\ee
In the classical information theory, this is known as Cramer-Rao bound and this bound was generalized to that in quantum information theory \cite{CR,Book}. Intuitively, this inequality means that if the density matrix $\rho_\theta$ changes more radically when we varies $\theta$, then the estimation gets easier.

If we apply this bound to a one parameter family of codimension two surfaces $\Sigma_u$ in our surface/state correspondence, we immediately obtain the inequality $\la (\delta u)^2\lb \geq \f{1}{8G^{(B)}_{uu}}$ from (\ref{cro}). The variance $\la (\delta u)^2\lb$ describes an error of the value of the coordinate $u$ in the gravitational spacetime $M_{d+2}$.

Especially, if we consider the AdS/CFT setup and identify $u$ with the radial coordinate as in
(\ref{padsmetm}), then we get the following bound
\be
\left\la \left(\f{\delta z}{z}\right)^2 \right\lb=\la (\delta u)^2\lb \geq \f{1}{8G^{(B)}_{uu}}\sim
\f{z^d}{N_{deg}\cdot V_d}\sim \f{G_N}{\mbox{A}(\Sigma_u)}, \label{rao}
\ee
where we introduced the standard radial coordinate $z=\ep e^{-u}$ of the Poincare AdS. Also
$\mbox{A}(\Sigma_u)=R^d V_d z^{-d}$ denotes the actual area of the surface
$\Sigma_u$. For the large $N$ gauge theories, we have $N_{deg}\sim N^2$ and thus the error gets suppressed in the large $N$ limit.\footnote{If we take $\Sigma_u$ to be the $d$ dimensional plane with $u$ fixed, then $V_d$ is already infinite. However, even if we choose $\Sigma_u$ to be a open manifold with a finite volume at the same value of $u$, we expect the same evaluation (\ref{rao}).} This seems to be consistent with the standard expectation in the AdS/CFT that classical geometries appear in the large $N$ limit. Moreover we can also interpret (\ref{rao}) in terms of effective entropy as follows
\be
\left\la \left(\f{\delta z}{z}\right)^2 \right\lb \gtrsim \f{1}{S_{\mbox{eff}}(\Sigma_u)}\sim \f{1}{\# \mbox{Links intersected with} \ \Sigma_u},
\ee
where in the final qualitative relation, we employed the holographic interpretation of tensor networks. It is suggestive to rewrite this inequality in the following way:
\be
\la \left(\delta \mbox{A}(\Sigma_u)\right)^2\lb \gtrsim G_N\cdot \mbox{A}(\Sigma_u),
\ee
in terms of the area $\mbox{A}(\Sigma_u)$ of $\Sigma_u$. It will be a very intriguing
future problem to understand the precise physical meaning of such a bound in the quantum estimation theory from the viewpoint of AdS/CFT correspondence.

\section{Several Examples}

Here we briefly study several explicit examples to see how our proposals work.

\subsection{Pure AdS}

Consider the AdS space in the Poincare coordinate given by the metric (\ref{padsmetm}).
If we take the surface $\Sigma_u$ to be the constant $u$ slice with the time $t(=x^0)$ fixed, then (\ref{gbcft}) agrees with the cMERA result for the dual CFT as we already mentioned in the previous section.

It is also intriguing to choose the one parameter family $\Sigma_{x^i}$ of surfaces by taking a constant $x^i$ slice, where $i$ can be one of $1,2,\ddd,d$, with the time fixed. In this case the proposed formula (\ref{Eincons}) leads to $G^{(B)}_{uu}=0$. This is consistent with the dual quantum state calculation because there is the translational invariance in $x^i$ direction and the quantum state, which is actually a mixed state, is invariant under the shift of $x^a$.

\subsection{AdS Black hole vs AdS Soliton}

The AdS black hole and AdS soliton are related to each other by a Wick-rotation and their Euclidean spaces include a common cigar-like geometry, whose polar coordinate to be defined to be $(r,\theta)$ \cite{Witten}. At $r=0$, the $\theta$ circle shrinks to zero size smoothly.

In the AdS black hole, the Euclidean time is taken in the $\theta$ direction. A one parameter family of surfaces $\Sigma_r$ is defined by fixing $r$ with $\theta=0$, which is a point in the cigar. Therefore it is not possible to contract $\Sigma_r$ to zero size by a smooth deformation. Our conjecture tells us that the corresponding state is a mixed state $\rho(\Sigma_r)$ as expected from the thermal nature of black hole. The von-Neuman entropy of $\rho(\Sigma_r)$ coincides with the black hole entropy $S_{BH}$ for any $r$, while we can have the matching for the effective entropy: $S_{\mbox{eff}}(\Sigma_r)=S_{BH}$ only for $r=0$.

On the other hand, in the AdS soliton, the time direction is not included in the cigar. We define $\Sigma_r$ by fixing the value of $r$ with the time fixed. Then $\Sigma$ is wrapped on the $\theta$ circle. However this circle can be smoothly contractible to zero size and thus we can conclude that the corresponding state is a pure state  $|\Phi(\Sigma_r)\lb$ as expected.

\subsection{Flat Spaces}

If we consider the flat spacetime $R^{1,d+1}$ and choose $u$ to be one of the cartesian coordinates:
\be
ds^2=du^2+dx_\mu dx^\mu,
\ee
then we immediately find that the right-hand side of the information metric (\ref{Eincons}) is simply vanishing. On the other hand, the translation symmetry in $u$ direction implies $|\Phi(\Sigma_{u+du})\lb=|\Phi(\Sigma_u)\lb$ and this clearly explains why the left-hand side of (\ref{Eincon}) is also vanishing. Note that $\Sigma_u$ is itself an extremal surface and therefore its entanglement entropy $S_A$ for a finite area part of $A$ satisfies the volume law and is equal to the effective entropy $S_{\mbox{eff}}(A)$. This shows that the quantum state dual to $\rho^{\Sigma_u}_A$ is non-locally entangled, while the entanglement between finitely separated regions is vanishing.\footnote{This can be understood clearly by compactifying $\Sigma_u$ e.g. regarding $\Sigma_u$ as a limit of a large spherical surface.} This is consistent with the analysis in \cite{LT,ShTa} of flat space holography. See \cite{max} also for an interesting analysis of entanglement entropy in flat space holography from a different viewpoint.

We can generalize this argument to the class of metric
\be
ds^2=du^2+g_{\mu\nu}(x)dx^\mu dx^\nu.
\ee
In this case we again have the translational symmetry in the $u$ direction, we find the information metric (\ref{Eincon}) should be trivial and this agrees with the dual description by quantum states.

\subsection{Discussion: De Sitter Spaces}

A much more non-trivial example will be de Sitter spaces. A holography for de Sitter spaces, called dS/CFT has been proposed in \cite{St} and it has been suggested that if gravity theories on de Sitter spaces can be dual to CFTs, then they are non-unitary CFTs \cite{MaldS} (see also \cite{AHS,ADH,Nar,Sat}).

A global metric of $d+2$ dimensional de Sitter space is given by
\be
ds^2=R^2(-dt^2+\cosh^2t\ d\Omega_{d+1}^2),
\ee
where $d\Omega_{d+1}^2$ is the metric of $S^{d+1}$ with the unit radius.
First let us define $\Sigma_{t}$ to be an equator of $S^{d+1}$ on the constant $t$ slice. At $t=0$, it is clear that the surface $\Sigma_{t=0}$, which is given by a $d$ dimensional round sphere $S^d$, is an extremal surface. Therefore we find
the saturation of subadditivity $S_{\mbox{eff}}(\Sigma_A)=S^{\Sigma_{t=0}}_A$ for any $\Sigma_A$ whose size is less than the half of $\Sigma_{t=0}$. From this fact, we can conclude that each point in $\Sigma_{t=0}$ is only entangled with its antipodal point and there is no other entanglement.
This property of quantum entanglement is far from that of CFT vacua. It is also interesting to note also that the total effective entropy $S_{\mbox{eff}}(\Sigma_{t=0})$ coincides with the de Sitter entropy as follows from the relation (\ref{effdim}).

On the other hand, for the surfaces $\Sigma_{t}$ at $t\neq 0$, we find its interpretation looks puzzling. It is obvious that its effective entropy grows exponentially
$S_{\mbox{eff}}(\Sigma_{t}) \propto \left(\cosh(t)\right)^d$. However, if we consider the entanglement entropy when we trace some part of $\Sigma_{t}$, it turns out that its corresponding space-like extremal surface does not always exist. For example, in the late time limit $t\gg 1$, only if we choose the size of subsystem $\Sigma_A$ to be as small as $O(e^{-t})$, we can find an extremal surface which computes to the holographic entanglement entropy $S^{\Sigma_{t}}_A$. One possibility to resolve this problem might be to pick up extremal surfaces by complexifying the spacetime coordinates as in \cite{Nar,Sat}. This trick leads to negative or complex valued entanglement entropy, which might be due to non-unitary nature of dual CFTs. Another possibility is to dismiss such surfaces without proper extremal surfaces and to focus on other class of surfaces. We would like to leave more studies for future problems.

\section{Conclusions}

In this paper, we proposed a new duality: surface/state correspondence, between codimension two space-like convex surfaces in gravity theory and quantum states in an infinitely large Hilbert space. Quantum states dual to topologically trivial closed surfaces are pure states, while those dual to open surfaces and topologically non-trivial closed surfaces are dual to mixed states. This proposal generalizes and refines the idea of holography and is highly motivated by the conjectured equivalence between AdS/CFT and tensor networks. It will be interesting to see more close connections between our formulation and the tensor network description. For example, our consideration of extremal surfaces for the calculation of entanglement entropy leads to the requirement of convexity and it will be important to understand its tensor network counterpart.
This looks related to the irreversibility of the coarse-graining operations.

Moreover, we studied some properties of these states, such as entanglement entropy, effective entropy and Fisher information metric. The entanglement entropy is given by the area of an extremal surface, which is a generalization of holographic entanglement entropy. The effective entropy is simply given by the area of the surface. The information metric is related to an integral of a square of extrinsic curvatures. Using our evaluation of the information metric, we applied the Cramer-Rao bound in quantum estimation theory and showed that the quantum fluctuations of estimated value of the radial coordinate of AdS space is suppressed in the large $N$ limit.
 It will be nice if we can find the precise form of the tensor $P$ in the expression of information metric from a direct calculation. Also it might be intriguing to study other quantities related to quantum entanglement such as the entanglement density introduced in \cite{NNT,BHRT} (see also \cite{LRSR} for a closely related work) and discuss how the Einstein equation appears in our formulation.

The most attractive feature of our proposal is that it does not rely on the existence of boundaries as opposed to the standard holography. In principle we can apply our proposal to understand holography for de-Sitter spaces as we briefly mentioned in this paper and we would like to come back to this problem in future publications. Also, in string theory we often encounter internal compact spaces such as $S^5$ in the type IIB background $AdS_5\times S^5$.
Even though interpretations of internal spaces have been discussed in \cite{MoST,KaUh} from the viewpoint of quantum entanglement, their realizations in tensor networks and in our surface/state correspondence are also remained as future problems.

\section*{Acknowledgements}

  We are grateful to Pawel Caputa, Veronika Hubeny, Esperanza Lopez, Mukund Rangamani and Joan Simon for useful discussions, and especially to Horacio Casini, Bartlomiej Czech and Xiao-Liang Qi for important comments and stimulating discussions on holographic interpretations of tensor networks. We would like to thank Masahiro Nozaki, Tokiro Numasawa, Shinsei Ryu, Noburo Shiba and Kento Watanabe for closely related collaborations and helpful discussions. TT also acknowledges the organizers and participants of the IFT workshop ``Entangle This: Space, Time and Matter'' for giving him an opportunity to present results of this paper and for useful feedbacks. TT is supported by JSPS Grant-in-Aid for Scientific Research (B) No.25287058 and JSPS Grant-in-Aid for Challenging Exploratory Research No.24654057 and also by World Premier International
Research Center Initiative (WPI Initiative) from the Japan Ministry
of Education, Culture, Sports, Science and Technology (MEXT).


\end{document}